\documentclass[authoryear,review,3p,times,sort&compress]{elsarticle}
\usepackage{color}
\usepackage{multirow,setspace,times,amssymb,amsmath,graphicx,color,subfigure,url}
\usepackage{dsfont,bm}
\usepackage{threeparttable}
\usepackage{rotating}
\usepackage{lscape,pdflscape}
\usepackage{dcolumn}
\usepackage{listings}
\usepackage{relsize}
\usepackage{booktabs}
\usepackage[colorlinks=true,linkcolor=blue,citecolor=blue,urlcolor=blue]{hyperref}
\usepackage[absolute,overlay]{textpos}
\usepackage{tikz}
\usepackage{caption}
\usepackage{pdflscape}
\usepackage{float}
\usepackage{afterpage}     
\usepackage{placeins}      
\usepackage{lipsum}        
\usepackage[capitalize]{cleveref} 
\usepackage{pdfpages}
\usepackage{overpic}
\usepackage{natbib}
\usepackage{dcolumn}
\graphicspath{{Figures/}}
\journal{Elsevier}

\begin{document}
	
	\begin{frontmatter}
		
		\title{Dynamic spillovers and investment strategies across artificial intelligence ETFs, artificial intelligence tokens, and green markets}

		\author[SUIBE]{Ying-Hui Shao}
		\author[SILC]{Yan-Hong Yang }
		\author[SB,RCE]{Han-Xian Zhou}
		\author[SB,RCE,SM]{Wei-Xing Zhou \corref{cor1}}
		\ead{wxzhou@ecust.edu.cn}
		\cortext[cor1]{Corresponding author. }

		\address[SUIBE]{School of Statistics and Information, Shanghai University of International Business and Economics, Shanghai 201620, China}
		\address[SILC]{SILC Business School, Shanghai University, Shanghai 201899, China}
		\address[SB]{School of Business, East China University of Science and Technology, Shanghai 200237, China}
		\address[RCE]{Research Center for Econophysics, East China University of Science and Technology, Shanghai 200237, China}
		\address[SM]{School of Mathematics, East China University of Science and Technology, Shanghai 200237, China}
		
		\begin{abstract}

			This paper investigates the risk spillovers among AI ETFs, AI tokens, and green markets using the $R^{2}$ decomposition method. We reveal several key insights. First, the overall transmission connectedness index (TCI) closely aligns with the contemporaneous TCI, while the lagged TCI is significantly lower. Second, AI ETFs and clean energy act as risk transmitters, whereas AI tokens and green bond function as risk receivers. Third,
			AI tokens are difficult to hedge and provide limited hedging ability compared to AI ETFs and green assets. Moreover, multivariate portfolios effectively reduce AI assets investment risk. Among them, the minimum correlation portfolio outperforms the minimum variance and minimum connectedness portfolios. 
			
		\end{abstract}
		\begin{keyword}
			AI ETFs \sep AI tokens \sep Green markets \sep  Risk spillovers \sep Portfolio management
		\end{keyword}
	\end{frontmatter}

	\section{Introduction}
	\label{Introduction}
	
	The onset of the Fourth Industrial Revolution has ushered human society into the era of artificial intelligence (AI). This transformation has not only revolutionized daily life, work practices, and communication methods but has also profoundly impacted education, healthcare, finance, and transportation \citep{Skilton-Hovsepian-2018}. Simultaneously, technological advancements have facilitated the evolution of financial markets, leading to the emergence of artificial intelligence related financial instruments. According to existing literature \citep{Yousaf-Youssef-Goodell-2024-JIntFinancMarkInstMoney,Yousaf-Ijaz-Umar-Li-2024-EnergyEcon,Yousaf-Ohikhuare-Li-Li-2024-EnergyEcon,Jareno-Yousaf-2023-IntRevFinancAnal}, AI-related assets constitute a class of financial instruments closely linked to artificial intelligence technologies and their broader industrial ecosystem. These assets can be broadly categorized into three groups: AI-themed exchange-traded funds (AI ETFs), publicly listed stocks of AI-related companies (AI-related stocks), and AI-integrated cryptocurrency assets (commonly referred to as AI tokens). AI ETFs allocate capital to companies in AI technology, robotics, cloud computing, and automation, offering diversified exposure to the AI sector \citep{Yousaf-Youssef-Goodell-2024-JIntFinancMarkInstMoney}. AI-related stocks include companies where AI is a core business component or holds a significant role in revenue and strategic positioning, such as NVIDIA and Google \citep{Yousaf-Ijaz-Umar-Li-2024-EnergyEcon,Yousaf-Ohikhuare-Li-Li-2024-EnergyEcon,Jareno-Yousaf-2023-IntRevFinancAnal}. And AI Tokens integrate AI and blockchain technology, primarily facilitating decentralized AI computing, data sharing, and smart contract execution to support distributed AI development \citep{Yousaf-Ijaz-Umar-Li-2024-EnergyEcon,Yousaf-Youssef-Goodell-2024-JIntFinancMarkInstMoney,Jareno-Yousaf-2023-IntRevFinancAnal}. 
	These innovative asset classes not only represent the forefront of technological progress but have also become 
novel investment instruments, attracting significant interest from institutional and individual investors  \citep{Yousaf-Youssef-Goodell-2024-JIntFinancMarkInstMoney}.

	The investment boom in the AI industry has driven the rapid expansion of computational infrastructure, providing strong momentum for technological innovation and productivity growth. However, this rapid development is heavily reliant on high-performance computing, which significantly increases energy consumption and raises environmental concerns \citep{Jones-2018-Nature, Vinuesa-Azizpour-Leite-Balaam-Dignum-Domisch-Fellander-Langhans-Tegmark-FusoNerini-2020-NatCommun}.	
	At the same time, AI plays a critical role in advancing the green economy by optimizing renewable energy integration, monitoring environmental indicators, enhancing energy efficiency, and supporting sustainability initiatives \citep{Vinuesa-Azizpour-Leite-Balaam-Dignum-Domisch-Fellander-Langhans-Tegmark-FusoNerini-2020-NatCommun, Entezari-Aslani-Zahedi-Noorollahi-2023-EnergyStrategRev, Nahar-2024-TechnolForecastSocChang}.
	
Amid the global push toward a low-carbon economy, financial markets are increasingly integrating environmental considerations into investment decisions and promoting green financing. As a result, green assets have gained prominence as key instruments in facilitating the transition to sustainability \citep{Tiwari-Abakah-Gabauer-Dwumfour-2022-GlobFinancJ, Henriques-Sadorsky-2024-GlobFinancJ, Tiwari-Abakah-Shao-Le-Gyamfi-2023-EnergyEcon}. Supported by favorable policies and strong growth prospects, these emerging markets are attracting rising investor interest, particularly in areas where green and AI-related technologies converge. 
	
	Given the complex relationship between artificial intelligence and green economic development, it is plausible that intricate financial interactions also exist between AI-related assets and green assets. Similarly, prior studies have documented dynamic linkages between green equities and technology stocks \citep{Sadorsky-2012-EnergyEcon, Henriques-Sadorsky-2024-GlobFinancJ, Tiwari-Abakah-Shao-Le-Gyamfi-2023-EnergyEcon}, and green assets have been shown to offer diversification benefits \citep{Tiwari-Abakah-Gabauer-Dwumfour-2022-GlobFinancJ,	  	Henriques-Sadorsky-2024-GlobFinancJ}.	
	Yet little is known about whether green assets can help diversify the risks of AI-related investments.

	Moreover, during periods of significant turmoil such as financial crises, the COVID-19 pandemic, and the Russia-Ukraine war, the interconnections between different financial markets tend to strengthen \citep{Henriques-Sadorsky-2024-GlobFinancJ,Demiralay-Gencer-Bayraci-2021-TechnolForecastSocChang,Tiwari-Abakah-Gabauer-Dwumfour-2022-GlobFinancJ,Zeng-Abedin-Zhou-Lu-2024-IntRevFinancAnal}. Despite their distinct mechanisms, these markets often experience similar impacts from such substantial external shocks. Under the current geopolitical tensions and ongoing economic uncertainties following the COVID-19 pandemic, investigating the information transmission between AI-related assets and green assets is both timely and of practical significance.

	Therefore, this study systematically investigates the risk spillover effects between AI assets and green assets via the $R^{2}$ decompostion method \citep{Balli-Balli-Dang-Gabauer-2023-FinancResLett,Naeem-Chatziantoniou-Gabauer-Karim-2024-IntRevFinancAnal}. Although AI-related assets encompass various forms, we specifically focus  on linkages among AI ETFs, AI Tokens and green assets. AI ETFs invest in multiple companies, providing a comprehensive representation of market trends in the AI sector, and capturing investor sentiment and capital flows effectively. In contrast, AI Tokens exhibit distinct market characteristics, such as high volatility, rapid evolution, and a decentralized digital financial structure. Frequent fluctuations in trading volumes, along with frequent token listings, delistings, and mergers, contribute to the market's dynamic nature. A recent example illustrating this volatility is Render’s rebranding from RNDR to RENDER and its migration from Ethereum to Solana.	
	Given these unique attributes, we further assess portfolio performance of AI ETFs and AI Tokens alongside green assets. We estimate the bilateral hedging ratios and hedge effectiveness of different assets  pairs. Additionally, we apply the the minimum variance portfolio (MVP)  \citep{Markowitz-1952-JFinanc}, and compare its performance with the minimum correlation portfolio (MCP) \citep{Christoffersen-Errunza-Jacobs-Jin-2014-IntJForecast}, and the minimum connectedness portfolio (MCoP) \citep{Broadstock-Chatziantoniou-Gabauer-2022}. 
	
	This paper has the following contributions: First, the development of green investment markets is essential for addressing climate change and promoting sustainable growth. At the same time, artificial intelligence has the potential to accelerate the global energy transition by improving efficiency and supporting innovative low-carbon solutions. Understanding the financial relationship between these two sectors can thus provide meaningful support for achieving sustainability goals.
	Second, this paper contributes to the literature by investigating risk spillovers between AI tokens, AI ETFs, and green assets. It reveals how market shocks may transmit across these asset classes and identifies both contemporaneous and lagged spillover effects. This approach enhances the understanding of dynamic interdependencies in increasingly integrated sustainable and tech-driven markets.
	Third, the study assesses the hedging effectiveness of AI-related assets and compares their performance with green assets under various investment strategies. These findings offer practical implications for portfolio diversification and asset allocation, while also helping regulators evaluate market stability through a risk transmission framework.

	The remainder of this paper is structured as follows. Section~\ref{Literature review} offers a review of the related literature.  Section~\ref{Data description} provides a description of the data sets, along with summary statistics. Section~\ref{Methodology} provides the methodological framework. Section~\ref{Dynamic connectedness analysis} presents the dynamic connectedness analysis. Section~\ref{Portfolio} discusses the portfolio and hedging strategies. Section~\ref{Conclusion} concludes the paper.

	\section{Literature review}
	\label{Literature review}
			
	The interconnections among AI-related assets and and other instruments have garnered significant attention, including stocks, bonds, commodities, cryptocurrencies, and electricity  \citep{Yousaf-Youssef-Goodell-2024-JIntFinancMarkInstMoney,Demiralay-Gencer-Bayraci-2021-TechnolForecastSocChang,Yousaf-Ohikhuare-Li-Li-2024-EnergyEcon,Abakah-Tiwari-Lee-NtowGyamfi-IntRevFinanc,Zeng-Abedin-Zhou-Lu-2024-IntRevFinancAnal,Sharma-Aggarwal-Dixit-Yadav-2023-Vision,Billah-2025-ResIntBusFinanc,Huynh-Hille-Nasir-2020-TechnolForecastSocChang,Ali-Al-Nassar-Khalid-Salloum-AnnOperRes,Asl-Roubaud-2024-FinancInnov,Abdullah-Sarker-Abakah-Tiwari-Rehman-2024-GlobFinancJ}. Overall, the empirical results demonstrate that the interlinkages vary across time–frequency domains and under different market conditions. \cite{Demiralay-Gencer-Bayraci-2021-TechnolForecastSocChang} find that the wavelet co-movements between AI and robotics stocks and other assets are time-scale dependent, with correlations increasing significantly during the COVID-19 pandemic. \cite{Yousaf-Ijaz-Umar-Li-2024-EnergyEcon} report that the connectedness between AI assets and fossil fuel markets varies across time-frequencies and quantiles, with short-term fluctuations driving shock transmission, while long-term dynamics can shift assets between net transmitter and receiver roles.
	 \cite{Abakah-Tiwari-Lee-NtowGyamfi-IntRevFinanc} examine the directional predictability between Bitcoin, fintech, and AI stocks. The results show that this predictability changes across different time lags and shows an oscillating pattern.

	A growing body of research explores the interdependence among different types of AI-based assets \citep{Yousaf-Youssef-Goodell-2024-JIntFinancMarkInstMoney,Billah-2025-ResIntBusFinanc,Yousaf-Ijaz-Umar-Li-2024-EnergyEcon,Jareno-Yousaf-2023-IntRevFinancAnal}. \cite{Billah-2025-ResIntBusFinanc} investigates the connections among AI-based financial assets, Sukuk, and Islamic equity indices using a Quantile VAR model. The study reveals distinct spillover patterns under extreme market conditions compared to normal periods, highlighting the uncertainty in the evolution of tail dependencies.  However, the correlation between AI stocks and tokens tends to be relatively low and may vary over time, increasing during periods of economic turmoil \citep{Jareno-Yousaf-2023-IntRevFinancAnal}. 	
	AI ETFs consistently act as net transmitters, dominating the system across market conditions, while AI tokens serve as weak net recipients at median quantiles, with all variables alternating between net transmitters and recipients over time \citep{Yousaf-Youssef-Goodell-2024-JIntFinancMarkInstMoney}. \cite{Yousaf-Ijaz-Umar-Li-2024-EnergyEcon} investigate volatility interconnections between AI tokens, AI stocks, and fossil fuel markets. They report that risk transmission patterns vary across AI-related assets, highlighting their heterogeneous roles as either transmitters or recipients. 
	
	Some scholars have examined the determinants of spillovers and interactions between AI-related assets and other asset classes. 
	\cite{Yousaf-Ohikhuare-Li-Li-2024-EnergyEcon} find that the impact of the  factors on spillover effects varies across different market conditions. They provide evidence that assets such as gold, Bitcoin, and bonds have a significant negative impact on the connectedness between the electricity market and the artificial intelligence market.	
	 \cite{Billah-2025-ResIntBusFinanc} examines how global factors influence asset spillovers among AI-based assets, Sukuk, and Islamic equity indices. The study finds that the impact of these factors on spillovers varies significantly depending on the prevailing market environment. 
	 \cite{Ali-Al-Nassar-Khalid-Salloum-AnnOperRes}	argue that economic policy uncertainty and U.S. stock market volatility increase the interconnectedness between AI and fintech stocks, while geopolitical risk and gold market fluctuations have a negative impact on their correlations.		
	 \cite{Abdullah-Sarker-Abakah-Tiwari-Rehman-2024-GlobFinancJ} find that global uncertainty factors significantly increase both upside and downside tail-risk connectedness among tech-industry tokens and equities.

	Several studies have pointed out that AI-focused assets are emerging as promising tools for diversification, providing investors with exposure to more balanced and diversified portfolios \citep{Jareno-Yousaf-2023-IntRevFinancAnal}. According to \cite{Billah-2025-ResIntBusFinanc}, Sukuk and AI-based token markets are key recipients of return shock spillovers, which has important implications for portfolio diversification and risk management strategies. 
	 \cite{Yousaf-Youssef-Goodell-2024-JIntFinancMarkInstMoney} find that under normal market conditions, AI tokens can serve as effective diversifiers within portfolios containing traditional assets. However, they point out that during periods of extreme market stress, neither AI tokens nor AI ETFs provide significant risk diversification benefits for other asset classes. Moreover, AI tokens have the potential to provide cost-effective hedging for traditional assets.
	\cite{Abdullah-Sarker-Abakah-Tiwari-Rehman-2024-GlobFinancJ} highlight substantial opportunities for portfolio diversification when combining tech-industry tokens with tech-industry equities.

	Moreover, there exits potential connections between AI- related assets and sustainable markets. The empirical findings indicate that the AI index consistently emits shocks to clean markets across all frequencies, while the S\&P Global Clean Energy Index functions as a net sender of volatility spillover across all market conditions and frequencies \citep{Zeng-Abedin-Zhou-Lu-2024-IntRevFinancAnal}.  Investors incorporating stocks of artificial intelligence and robotics companies, cryptocurrencies, and green bonds into their portfolios should be aware that certain investment risks persist \citep{Huynh-Hille-Nasir-2020-TechnolForecastSocChang}.

	In summary, existing research has extensively explored the correlations between AI-related assets and other financial assets, with some studies further investigating the underlying factors influencing these relationships. However, there remains a lack of systematic research on the correlation and interdependence between AI assets and green assets. 
	Furthermore, limited research has explored the role of sustainability-linked financial instruments in hedging risks and optimizing portfolio performance in relation to AI assets. Moreover, AI tokens, as a unique type of crypto asset, have introduced significant changes in global financial dynamics. 
	Hence, our study contributes to the literature on spillovers between innovative AI-based assets and green markets. We emphasize the potential diversification benefits that AI-based assets can offer in portfolios and highlight their influence on overall portfolio performance.

	\section{Data description and summary statistics}
	\label{Data description}
	
	Throughout this research, we examine three categories of highly capitalized or actively traded assets: Firstly, AI ETFs, specifically iShares Robotics and Artificial Intelligence ETF (IRBO) and First Trust Nasdaq Artificial Intelligence and Robotics ETF (ROBT), which are representative assets in the artificial intelligence sector, with data sourced from Investing.com. 
	
	Secondly, AI-based tokens, including Render (RENDER), NEAR Protocol (NEAR), Internet Computer (ICP), Filecoin (FIL), and Artificial Superintelligence Alliance (FET), obtained from CoinGecko. These tokens represent diverse applications within the blockchain and AI ecosystems. The selection of tokens in this study is based on market capitalization, trading volume, establishment date, and rankings as reported by CoinMarketCap. Tokens with lower rankings, insufficient market capitalization, and trading volume, or a relatively short history were excluded from consideration.	
	Among these tokens, FET was formed through the merger of Fetch.ai, SingularityNET, and Ocean Protocol, and was later rebranded as the Artificial Superintelligence token. In this study, it is referred to as FET for simplicity. 
    RENDER is a platform focused on decentralized graphics rendering. As of this analysis, it has transitioned from its previous name, Render (RNDR), to Render (RENDER). For consistency, we use ``RENDER'' as its shorthand throughout this article, with Render as the full name.	
	NEAR is dedicated to developing a scalable blockchain platform. ICP serves as the native token of the Internet Computer Protocol. 
	FIL functions as the utility token for the Filecoin network, and FET is associated with an AI-powered decentralized machine learning network. 
	
	Thirdly, green assets, comprising the S\&P Green Bond (SPGB) and S\&P Global Clean Energy Transition Index (SPGTCED), were collected from the S\&P Global website. Due to the launch date of the ICP being May 10, 2021, and after excluding extreme fluctuations, the time period for this study spans from May 25, 2021, to January 31, 2025. The analysis utilizes daily data, with a total of 927 observations. All asset returns are calculated using logarithmic returns.

	\begin{table}[htp]
		\centering
			\caption{Descriptive statistics.}
			\newcolumntype{d}[1]{D{.}{.}{#1}}
			\label{Tb:Statistics}
\begin{tabular*}{\textwidth}{@{\extracolsep{\fill}}l *{4}{d{2.3}} d{5.6} d{2.6} @{}}
				\toprule
				& \multicolumn{1}{c}{\textbf{Mean ($\times 10^2$) }} 
				& \multicolumn{1}{c}{\textbf{Variance ($\times 10^2$) }} 
				& \multicolumn{1}{c}{\textbf{Skewnes}} 
				& \multicolumn{1}{c}{\textbf{Ex.Kurtosis}}
				& \multicolumn{1}{c}{\textbf{JB}} 
				& \multicolumn{1}{c}{\textbf{ERS}} \\
				
				\midrule
				IRBO & 0.005 & 0.028 & -0.003 & 1.717^{***} & 113.810^{***} & -12.922^{***}  \\
				ROBT & 0.002 & 0.024 & 0.069 & 1.264^{***} & 62.427^{***} & -14.157^{***} \\
				FET & 0.494 & 0.677 & 1.252^{***} & 7.151^{***} & 2217.373^{***} & -1.276^{***} \\
				FIL & -0.054 & 0.474 & 1.738^{***} & 17.811^{***} & 12719.882^{***} & -2.573^{***}  \\
				ICP & -0.054 & 0.504 & 1.352^{***} & 10.751^{***} & 4746.375^{***} & -9.715^{***} \\
				NEAR & 0.330 & 0.558 & 0.748^{***} & 4.623^{***} & 911.714^{***} & -2.164^{***} \\
				RENDER & 0.642 & 0.801 & 1.130^{***} & 4.687^{***} & 1045.970^{***} & -1.498^{***} \\
				SPGB &  -0.019& 0.023 & 0.247^{***} & 1.825^{***} & 138.021^{***} & -6.615^{***} \\
				SPGTCED & -0.061 & 0.025 & 0.454^{***} & 2.307^{***} & 237.196^{***} & -14.460 ^{***} \\
				\bottomrule
			\end{tabular*}  
			\begin{tablenotes}
				\footnotesize
				\item Note: * , **, *** indicate significance at the 10\%, 5\%, and 1\% levels, respectively.
			\end{tablenotes}
	\end{table}

	Table~\ref{Tb:Statistics} presents the descriptive statistics and preliminary tests for the AI and green assets. The mean returns vary significantly across assets, with FET and RENDER exhibiting the highest average returns, while FIL, ICP, SPGB, and SPGTCED display negative means. Variance suggests that AI tokens, particularly RENDER, exhibit higher volatility compared to AI ETFs and green assets. Skewness values indicate asymmetry in return distributions. While most assets exhibit positive skewness, IRBO shows slight letftward skew. Excess kurtosis values are significantly positive for all assets, implying leptokurtic distributions. The Jarque-Bera (JB) test strongly rejects normality for all assets at the 1\% level. Additionally, the Elliott-Rothenberg-Stock (ERS) test results indicate strong stationarity for all series, as evidenced by significantly negative values. These findings suggest non-normal, highly volatile return distributions, particularly among AI tokens.
	
	\begin{figure}[htp]
		\centering
		\includegraphics[width=0.6\linewidth]{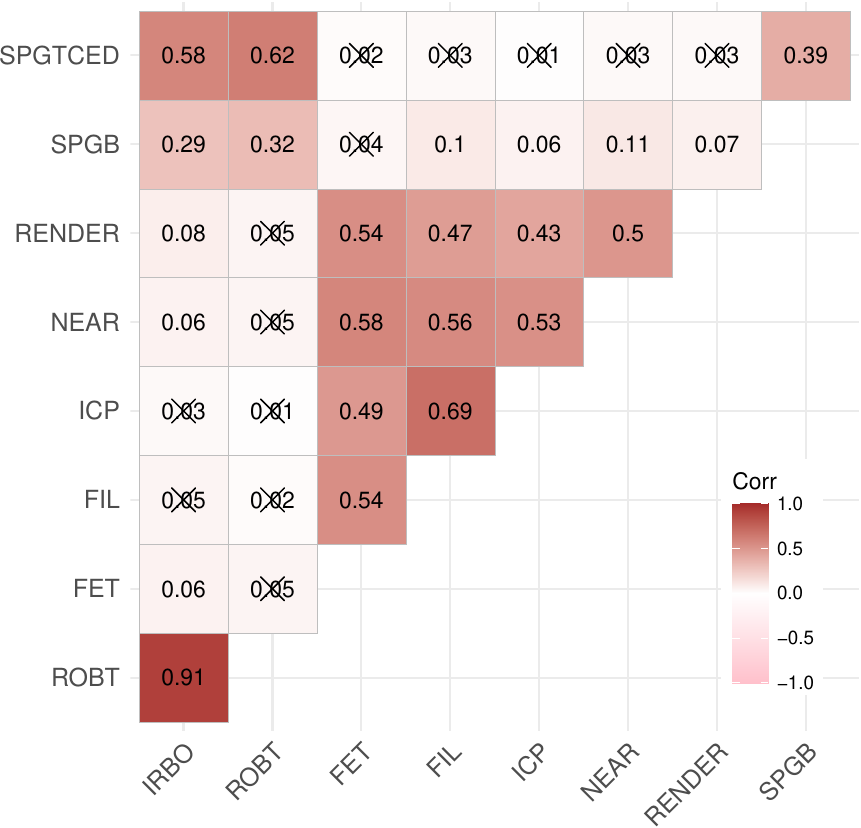} 
		\caption{Pairwise correlation heatmap. Notes: Coefficients that do not meet the 10\% significance level are shown as crossed-out values.}
		\label{Fig:Correlation}
	\end{figure}
	
	Fig.~\ref{Fig:Correlation} presents a pairwise correlation heatmap for the selected assets, where the shading intensity reflects correlation strength. Darker red shades indicate higher positive correlations, while lighter shades correspond to weaker correlations. Correlation coefficients that do not meet the 10\% significance level are marked with crossed-out values. The strongest correlation is observed between IRBO and ROBT (0.91), indicating a strong co-movement between these two AI-related ETFs. Among AI tokens, ICP and FIL (0.69) exhibit a relatively strong correlation, suggesting potential co-movement in their market dynamics, while NEAR and RENDER also display moderate correlations with other tokens, with values ranging from 0.43 to 0.58. 
	
	In contrast, the correlations between SPGB (green bonds) and AI assets remain relatively weak, with values mostly below 0.3, suggesting limited interaction between these asset classes. However, SPGTCED (another green asset) shows a stronger correlation with AI ETFs, reaching 0.62 with ROBT and 0.58 with IRBO, suggesting some level of connectedness between sustainable investment themes and AI-driven assets. Notably, The correlation coefficients between ROBT and AI tokens, as well as those between clean energy assets and cryptos, largely fail to pass the significance test.
	
	In conclusion, the results indicate strong intra-category correlations. Green assets exhibit moderate correlations with AI ETFs. However, AI tokens have weak or insignificant correlations with other asset classes.

	\section{Methodology}
	\label{Methodology}
	
	\subsection{$R^{2}$ decomposition approach}
	In this study, we utilize the innovative $R^2$ decomposition technique to examine the dynamic linkages among AI-based ETFs, Tokens, and green markets over different time periods \citep{Balli-Balli-Dang-Gabauer-2023-FinancResLett,Naeem-Chatziantoniou-Gabauer-Karim-2024-IntRevFinancAnal}. This approach efficiently disaggregates relevance into contemporaneous and lagged components while ensuring standardized data processing. By overcoming the limitations of prior research, which predominantly concentrated on contemporaneous spillover effects, this method broadens the scope of investigation \citep{Balli-Balli-Dang-Gabauer-2023-FinancResLett}. The total connectedness index (TCI) is computed as the average $R^2$ across $k$ multivariate linear regressions:
	\begin{equation}
	TCI=\frac{1}{K}\sum_{k=1}^{K}R_{k}^{2}=\left(\frac{1}{K}\sum_{k=1}^{K}\sum_{i=1}^{K}\bm{R}_{C,k,i}^{2,d}\right)+\left(\frac{1}{K}\sum_{k=1}^{K}\sum_{i=1}^{K}\bm{R}_{L,k,i}^{2,d}\right)=TCI^{C}+TCI^{L},
	\end{equation}
	where $ \bm{R}^{2,d}_C $ and $ \bm{R}^{2,d}_L $ refer to contemporaneous and lagged spillover effects, respectively. The total directional spillovers, both to and from other markets, as well as the net spillover, are calculated as follows:
	\begin{align}
	TO_{i} &= \sum_{k=1}^{K}\bm{R}_{C,k,i}^{2,d} + \sum_{k=1}^{K}\bm{R}_{L,k,i}^{2,d} = TO_i^C + TO_i^L ,
	\end{align}
	\begin{align}
	FROM_{i} &= \sum_{k=1}^{K}\bm{R}_{C,i,k}^{2,d} + \sum_{k=1}^{K}\bm{R}_{L,i,k}^{2,d} = FROM_i^C + FROM_i^L, 
	\end{align}
	\begin{align}
	NET_i &= TO_i - FROM_i \\
	&= (TO_i^C + TO_i^L) - (FROM_i^C + FROM_i^L) \\
	&= (TO_i^C - FROM_i^C) + (TO_i^L - FROM_i^L) \\
	&= NET^{C}_i + NET^{L}_i.
	\end{align}
	When $NET_i > 0$ (or $NET_i < 0$), the market $i$ is identified as a net contributor (or receiver) of shocks. Furthermore, net pairwise directional connectedness (NPDC) is calculated in a similar manner to NET, but on a pairwise level \citep{Balli-Balli-Dang-Gabauer-2023-FinancResLett}.

	\subsection{Portfolio analysis}
	
	\subsubsection{Bilateral hedge ratios and portfolio weights}
	
	According to 
	\cite{Kroner-Sultan-1993-JFinancQuantAnal}, the dynamic hedge ratio is expressed as:
	\begin{equation}
	\beta_{ij,t} = \frac{\Sigma_{ij,t}}{\Sigma_{jj,t}},
	\end{equation}
	where $\Sigma_{ij,t}$ and $\Sigma_{jj,t}$ represents the conditional covariance and the conditional variance at time $t$.
	
	Following 
	\cite{Kroner-Ng-1998-RevFinancStud}, the optimal bilateral portfolio weights are given by:
	\begin{equation}
	w_{ij,t} = \frac{\Sigma_{ii,t} - \Sigma_{ij,t}}{\Sigma_{ii,t} - 2\Sigma_{ij,t} + \Sigma_{jj,t}},
	\end{equation}
	with
	\begin{equation}
	w_{ij,t} =
	\begin{cases} 
	0, & \text{if } w_{ij,t} < 0 \\
	w_{ij,t}, & \text{if } 0 \leq w_{ij,t} \leq 1 ,\\
	1, & \text{if } w_{ij,t} > 1
	\end{cases}
	\end{equation}
	where $w_{ij,t}$ denotes the proportion of series $i$ in a $1$-dollar portfolio consisting of series $i$ and $j$ at time $t$. Consequently, $1 - w_{ij,t}$ represents the weight of series $j$ within the same portfolio.
	
	\subsubsection{Minimum variance portfolio}
	
	The minimum variance portfolio (MVP) forms lowest possible risk for a set of assets \citep{Markowitz-1952-JFinanc}, which is widely used in portfolio management to construct a low-volatility portfolio. 
	The optimal weights for MVP are given by:
	\begin{equation}
	\mathbf{w}_{\Sigma_t} = \frac{\Sigma_t^{-1} I}{I^T \Sigma_t^{-1} I},
	\end{equation}
	where $\mathbf{w}_{\Sigma_t}$ represents the portfolio weight vector, $I$ is a $K$-dimensional vector of ones, and $\Sigma_t$ denotes the $K \times K$ conditional variance-covariance matrix at time $t$.
	
	\subsubsection{Minimum correlation portfolio}
	
	The minimum correlation portfolio (MCP) method is another widely used portfolio construction approach, introduced by 
	 \cite{Christoffersen-Errunza-Jacobs-Jin-2014-IntJForecast}. This method constructs portfolios by minimizing conditional correlation. The portfolio weights under MCP are computed as follows:
	\begin{equation}
	R_t = \text{diag}(\Sigma_t)^{-0.5} H_t \text{diag}(\Sigma_t)^{-0.5},
	\end{equation}
	where $R_t$ is the conditional correlation matrix for the $K \times K$ dimensional portfolio at time $t$. The weights in MCP are calculated as:
	\begin{equation}
	\omega_{R_t} = \frac{R_t^{-1} I}{I^T R_t^{-1} I},
	\end{equation}
	where $\omega_{R_t}$ is the weight vector for the $K \times 1$ dimensional portfolio, and $I$ represents the $K$-dimensional vector of ones.
	
	\subsubsection{Minimum connectedness portfolio}
	
	The minimum connectedness portfolio (MCoP) approach constructs a portfolio by minimizing the pairwise connectedness of indices \citep{Broadstock-Chatziantoniou-Gabauer-2022}.  The asset weights are computed as follows:
	\begin{equation}
	\omega_{C_t} = \frac{PCI_t^{-1} I}{I PCI_t^{-1} I},
	\end{equation}
	where $ \omega_{C_t} $ represents the weight of an asset in the portfolio, $ PCI_t $ denotes the pairwise connectedness index matrix at time \( t \), and \( I \) is the identity matrix.
	
	\subsubsection{Portfolio analysis}
	
	To evaluate portfolio performance, two key metrics are used: the Sharpe ratio (SR) \citep{Sharpe-1966-JBus} 
	and hedging effectiveness (HE) \citep{Ederington-1979-JFinanc}.
	
	The Sharpe ratio is computed as:
	\begin{equation}
	SR = \frac{\bar{r}_p}{\sqrt{\text{var}(r_p)}},
	\end{equation}
	where \( r_p \) represents portfolio returns assuming a risk-free rate of zero. A higher SR indicates superior returns relative to portfolio risk, facilitating comparison across portfolios with similar volatility.
	
	The hedging effectiveness (HE) measures the percentage risk reduction achieved by the portfolio compared to investing in a single asset \( i \) \citep{Ederington-1979-JFinanc}. It is given by:
	\begin{equation}
	HE_i = 1 - \frac{\text{var}(r_p)}{\text{var}(r_i)},
	\end{equation}
	where \( \text{var}(r_p) \) denotes the portfolio variance and \( \text{var}(r_i) \) represents the variance of asset \( i \). A higher HE value signifies greater risk reduction.

	%
	
	\section{Dynamic connectedness analysis}
	\label{Dynamic connectedness analysis}

	\subsection{Averaged spillover effects}

\begin{figure}[t!]
		\centering
		\includegraphics[width=0.99\linewidth]{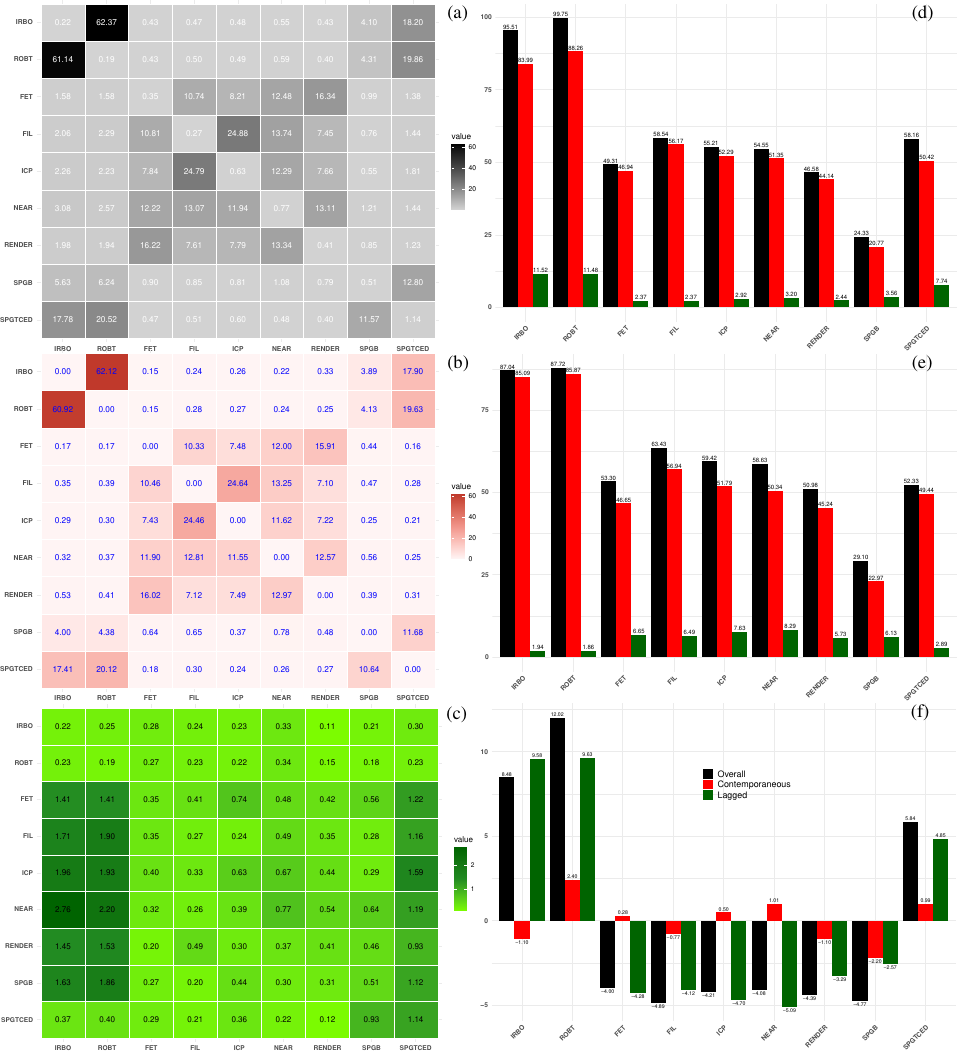}
		\caption{Averaged dynamic spillover indices. Notes: In the heatmap, the $ (i, j) $ element represents the magnitude of the spillover effect from asset $ j $ to asset $ i $. Panels (a) through (f) correspond to overall, contemporaneous, lagged, TO, FROM, and NET connectedness, respectively.}
		\label{Fig:Static:Connectedness}
	\end{figure}

	 Fig.~\ref{Fig:Static:Connectedness}~(a-c) present the heatmaps of the average overall, contemporaneous, and lagged connectness. Darker colors in the figure indicate higher levels of risk spillover. The $(i, j)$ element represents the average spillover from asset $j$ to asset $i$, while the $(j, i)$ element denotes the average spillover from asset $i$ to asset $j$. The difference between these values reflects the net average spillover from asset $j$ to asset $i$. 
	
	A detailed inspection of Fig.~\ref{Fig:Static:Connectedness}~(a-b) indicates that the overall and contemporaneous average spillover effects exhibit similar patterns. First, within the same asset category, spillover effects between assets tend to be stronger. For instance, the overall spillovers from ROBT to IRBO and from IRBO to ROBT are 62.37 and 61.14, respectively, indicating the strongest risk transmission between these two assets. This result is consistent with their highest positive correlation, as shown in Fig.~\ref{Fig:Correlation}. Green assets and AI tokens exhibit a similar pattern. 
	Second, the level of information transmission among different types of AI assets remains relatively moderate, mirroring the weak correlation between AI stocks and AI tokens \citep{Jareno-Yousaf-2023-IntRevFinancAnal}. 
	Third, spillover effects between clean energy assets and AI ETFs are also relatively strong, indicating notable risk transmission between these asset classes. Moreover, within the AI token market, FIL and ICP exhibit the strongest mutual spillover transmission.
	
	Furthermore, AI ETFs act as major net spillover transmitters to both green bonds and AI tokens, a finding that aligns with the results reported by 
	\cite{Yousaf-Youssef-Goodell-2024-JIntFinancMarkInstMoney} and 
	\cite{Huynh-Hille-Nasir-2020-TechnolForecastSocChang}. A more detailed examination reveals that ROBT transmits net spillovers to all assets, including IRBO and clean energy. This result is consistent with work  of 
	\cite{Zeng-Abedin-Zhou-Lu-2024-IntRevFinancAnal}, which highlights that the NASDAQ CTA Artificial Intelligence \& Robotics Index, tracked by ROBT, also transmits risk to clean energy. Additionally, the spillover dynamics of AI tokens exhibit strong similarities, potentially driven by the herding effect in the artificial intelligence and big data token markets. Investors appear to be influenced by collective behavior, leading to coordinated market movements that align with prevailing market consensus \citep{Xu-Ali-Naveed-ResIntBusFinanc}.
	
	However, a closer look at  Fig.~\ref{Fig:Static:Connectedness} (a-c) reveals distinctions among the averaged overall, contemporaneous, and lagged spillover effects. Specifically, in the overall analysis,
	clean energy transmits net spillovers to all assets, except ROBT. In the contemporaneous period, all assets transmit net spillovers to green bond. In the lagged period, clean energy transmits spillovers to tokens and bonds.

	Observing Fig.~\ref{Fig:Static:Connectedness} (d-f),  the three bar charts on the right illustrate the averaged spillover effects among different assets, with each panel representing overall (black), contemporaneous (red), and lagged (green) connectedness. Fig.~\ref{Fig:Static:Connectedness} (d) displays the TO connectedness, indicating how much each asset contributes to risk spillovers within the system. AI ETFs (IRBO and ROBT) and clean energy have the highest TO connectenss, indicating that they are the primary risk transmitters, with most spillover effects occurring contemporaneously rather than with a lag. Fig.~\ref{Fig:Static:Connectedness}  (e) presents FROM connectedness, measuring the extent to which each asset receives risk from others. AI ETFs and clean energy continue to exhibit the most significant FROM values, suggesting that they not only transmit risk but also absorb it from other assets. Lagged effects remain relatively small, indicating that risk transmission is predominantly immediate. Fig.~\ref{Fig:Static:Connectedness}  (f) represents NET connectedness, calculated as the difference between TO and FROM spillovers. AI ETFs have the highest net spillover values, reinforcing their role as key risk transmitters, while AI tokens and green bond generally exhibit negative net spillover values, confirming their role as risk recipients. Notably, SPGTCED has a positive net spillover value, suggesting that it still transmits a certain level of risk to other assets.

	Overall, Fig.~\ref{Fig:Static:Connectedness} (d-f) suggest that AI ETFs and clean energy serve as the primary sources of risk transmission, whereas AI tokens and green bonds mainly act as risk recipients, with NET spillover effects occurring over extended periods. The empirical results are in line with work of 
	\cite{Billah-2025-ResIntBusFinanc}, 
	\cite{Yousaf-Ijaz-Umar-Li-2024-EnergyEcon}, and 
	\cite{Yousaf-Youssef-Goodell-2024-JIntFinancMarkInstMoney}.

	\subsection{Dynamic total connectedness measures}

\begin{figure}[htp]
		\centering
		\includegraphics[width=0.4\linewidth]{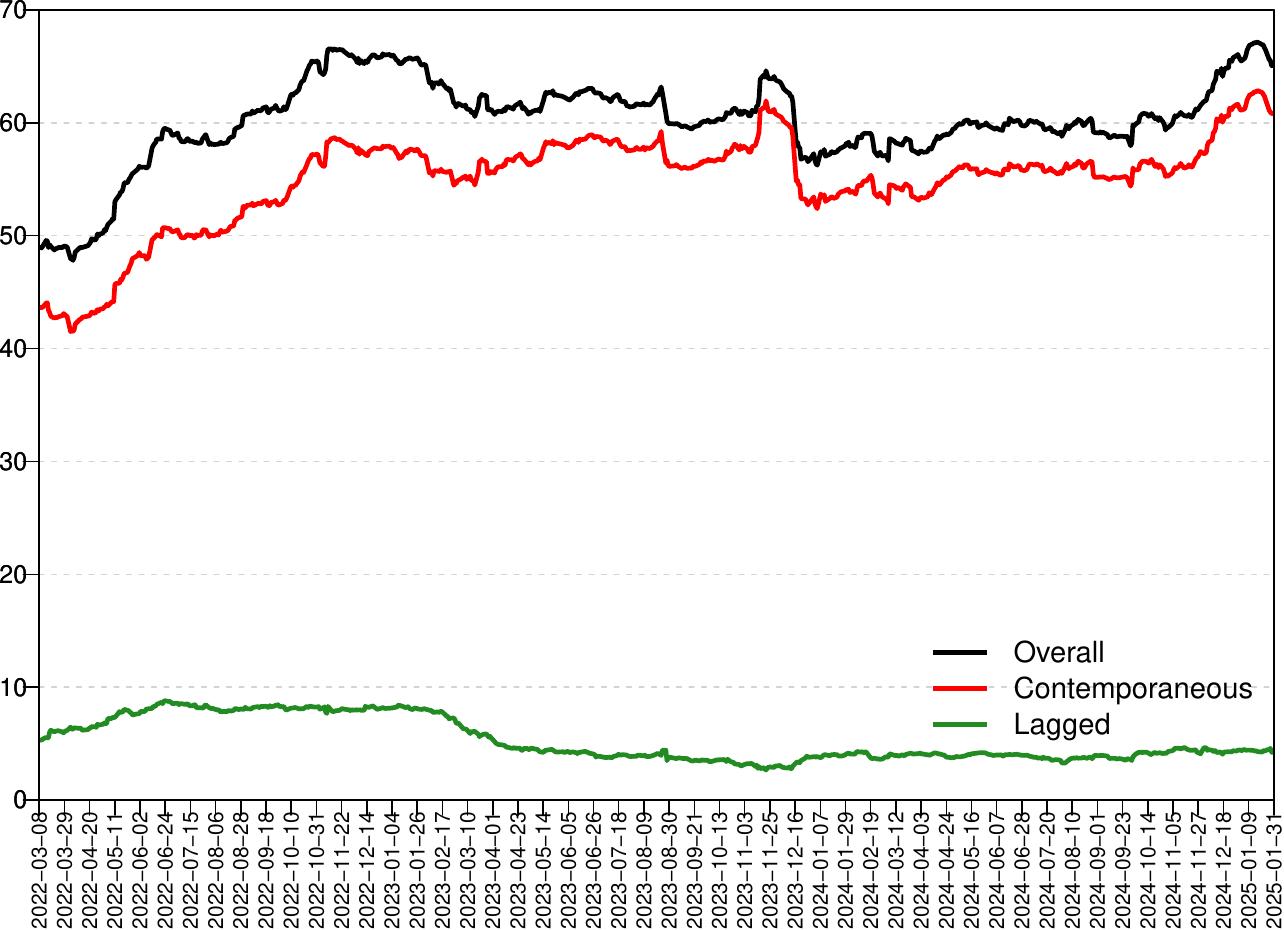}
		\caption{Evolution of total connectedness. }
		\label{Fig:TCI}
	\end{figure}

	Fig.~\ref{Fig:TCI} presents the dynamic total connectedness,  where both overall and contemporaneous connectedness exhibit a similar upward trend. The first peak occurs in mid-December 2022, reaching approximately 64\%, followed by a decline. Throughout 2023, some fluctuations are observed. The second major peak occurs in mid-January 2025, reaching approximately 66\%. In contrast, lagged TCI remains consistently low, fluctuating below 10\% throughout the entire period.
	
	In summary, the strong synchronization between overall and contemporaneous connectedness suggests that risk spillovers primarily occur in real-time rather than with a lag. The consistently low lagged connectedness indicates that market information transmission does not experience significant delays. 
	
	\subsection{NET total directional connectedness}

\begin{figure}[H]
		\centering
		\includegraphics[width=0.8\linewidth]{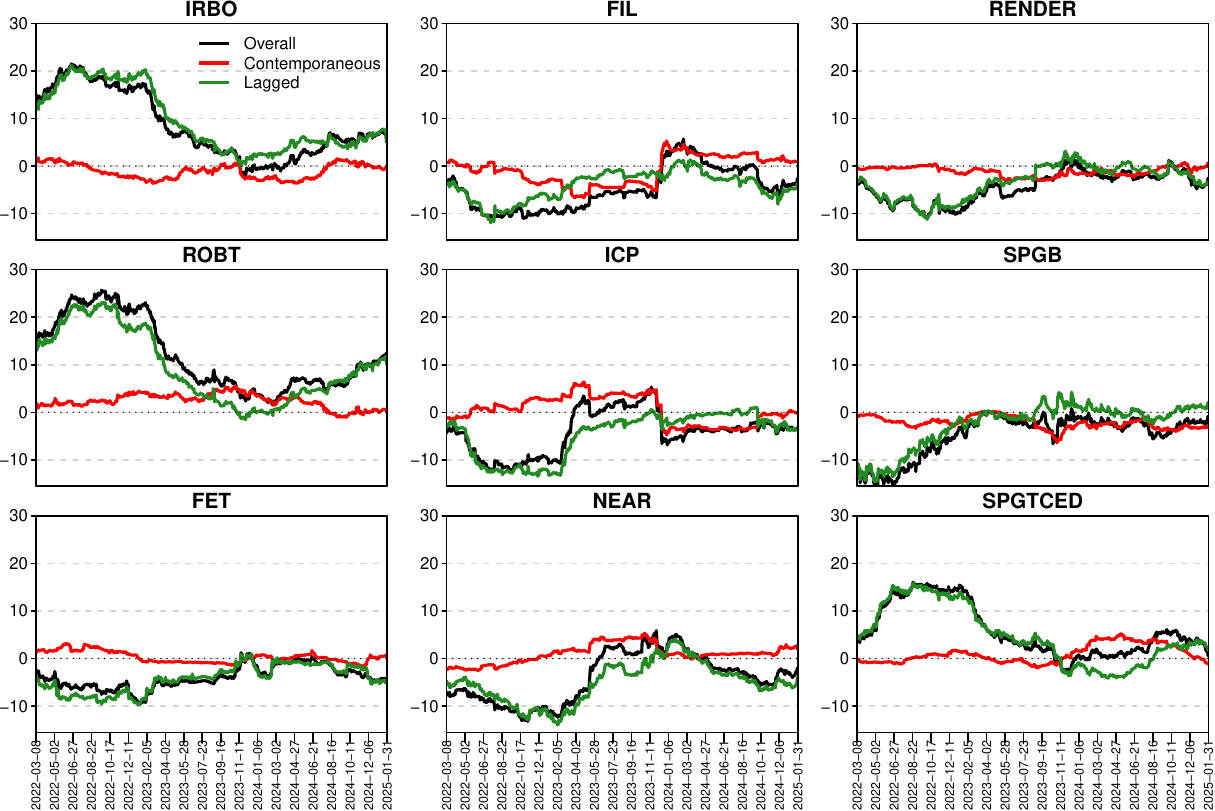}
		\caption{NET total directional connectedness.}
		\label{Fig:NET}
	\end{figure}

	Fig.~\ref{Fig:NET} illustrates the time-varying directional connectedness of AI and green assets. In line with Fig.~\ref{Fig:Static:Connectedness}~(f), it can be observed that the overall spillover effects and lagged spillover effects exhibit a closely aligned trend across various asset classes. In contrast, the net contemporaneous spillover effects  show relatively weaker fluctuations, with certain assets experiencing significant increases in spillover effects during specific periods.

	Moreover, the AI ETFs and clean energy assets exhibit a similar trend, both reaching a peak around October 2022, followed by a decline, and then an upward trajectory starting in 2024. Their overall spillovers remain predominantly positive throughout the period, consistently serving as the strongest contributors to systemic risk, highlighting their role as key drivers of market fluctuations. 
	\cite{Zeng-Abedin-Zhou-Lu-2024-IntRevFinancAnal} and we both conclude that clean energy acts as a net transmitter. Additionally, this study finds that AI funds also serves as a net transmitter, which is consistent with 
	\cite{Yousaf-Youssef-Goodell-2024-JIntFinancMarkInstMoney}. In line with \cite{Billah-2025-ResIntBusFinanc}'s 
	 findings, AI tokens likewise act as recipients of shocks within the system.
	
	In addition, AI tokens and SPGB maintain relatively low levels of spillover dynamics, which remain negative for most of the period, except between mid-2023 and mid-2024. Their overall spillover behavior is almost the opposite of AI ETFs and clean energy assets, indicating that they generally act as recipients of shocks within the system. This is particularly evident during periods of heightened market volatility, where their net spillover inflows become more pronounced. However, it is important to note that the spillover roles of these assets evolve dynamically over time, and in certain periods, they may also serve as risk transmitters.
	
	Generally speaking, Fig.~\ref{Fig:NET} reflects the complex interrelationships between the AI and green asset markets. AI ETFs and clean energy assets exhibit strong similarities, whereas AI tokens and the green bond market follow distinct spillover effect patterns. This suggests that different types of AI and green assets interact differently with systemic risk, highlighting the necessity of further exploration into their interconnectedness under varying market conditions.

	\subsection{Net pairwise directional connectedness measures}
	
	\begin{figure}[htp]
		\centering
		\includegraphics[width=0.95\linewidth]{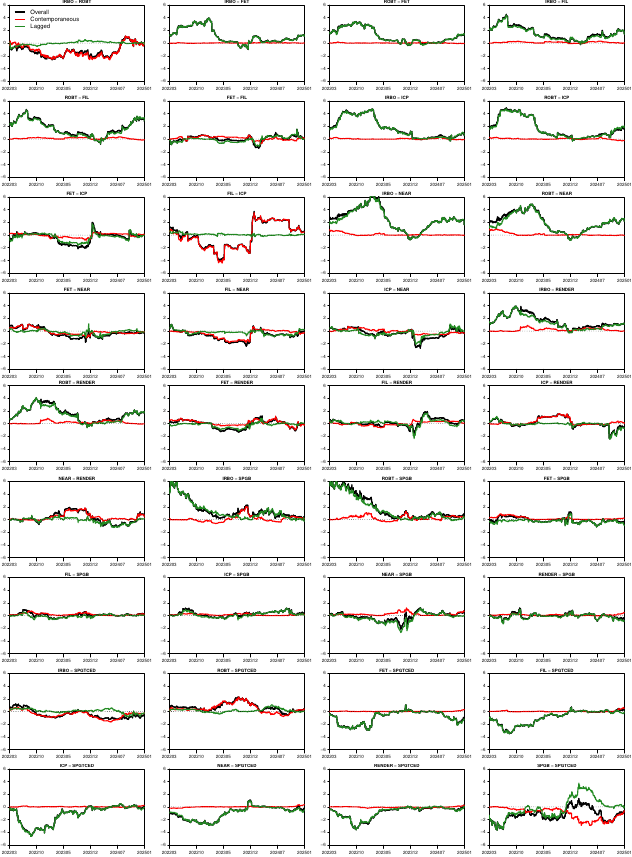}
		\caption{Net pairwise directional connectedness.}
		\label{Fig:NPDC}
	\end{figure}
	
	Fig.~\ref{Fig:NPDC} illustrates the net pairwise directional spillover. The dynamics of risk transmission from these two ETFs to other assets are highly similar, and IRBO primarily receives risk from ROBT. Aligning with the findings of 
	\cite{Huynh-Hille-Nasir-2020-TechnolForecastSocChang}, AI funds exhibits high overall spillovers. Moreover, the spillover effects between AI ETFs and AI tokens exhibit a similar pattern, generally increasing throughout the sample period until reaching a peak, followed by a decline, reaching their lowest levels between late 2023 and the first quarter of 2024, before rising again.

	Besides that, there exists significant spillovers from AI based ETFs to green bond, in line with Fig.~\ref {Fig:Static:Connectedness}. The connectedness declines until the end of 2023 and then stabilizes, fluctuating around zero. The spillover from AI ETFs to green bonds is generally stronger than the spillover from AI ETFs to clean energy. In addition, all AI tokens receive spillovers from clean energy with similar dynamics. It first reaches its maximum value in the second half of 2022, then declines and fluctuates around zero. NEAR and clean energy are exceptions, as their connectedness reaches its strongest level in the first quarter of 2023. 
	Overall, assets of the same type generally alternate between the roles of transmitter and receiver. AI ETFs and clean energy mainly acts as contributors of system shocks to other assets, which accords with Fig.~\ref{Fig:NET}.

	\subsection{Net pairwise network}

\begin{figure}[H]
		\centering
		\includegraphics[width=0.95\linewidth]{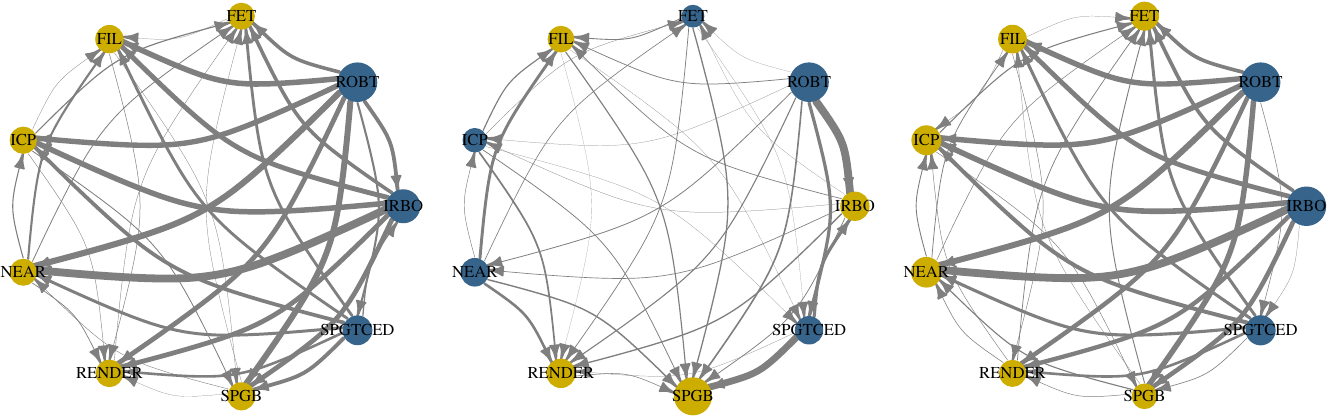}
		\caption{Net pairwise directional spillover networks.The three panels, from left to right, illustrate overall, contemporaneous, and lagged dependencies.}
		\label{Fig:Network}
	\end{figure}

	Fig.~\ref{Fig:Network} depicts the net pairwise directional spillover networks, filtering edges below 0.05 to highlight major spillover effects. Blue nodes represent transmitters, while yellow nodes denote receivers. The thickness of edges indicates the strength of spillover effects. In line with the results documented by 
	\cite{Tiwari-Abakah-Gabauer-Dwumfour-2022-GlobFinancJ}, our study also identifies a risk spillover effect from clean energy to green bonds. Again, the AI ETFs are the primary shock transmitters to other assets. Connectedness among AI tokens are weaker than their links with AI ETFs and green assets.
	The overall and lagged networks suggests the strong  spillovers from AI ETFs and clean energy to other assets. ROBT is the largest transmitter, followed by IRBO, and finally clean energy. 
	\cite{Yousaf-Youssef-Goodell-2024-JIntFinancMarkInstMoney} also report that AI ETFs (tokens) are strong (weakest) net transmitters (recipients) during normal market conditions. AI tokens and green bond receive relatively similar level of shocks from other markets. 
	
	However, the contemporaneous network indicates that IRBO, FIL, RENDER, and SPGB acts as receivers. Green bond receives most shocks from other markets. Further, FET, ICP, and NEAR transmit risk, although the level of shock remains relatively low.
	
	\subsection{Robustness test}	
	
\begin{figure}[htp]
		\centering
		\includegraphics[width=0.65\linewidth]{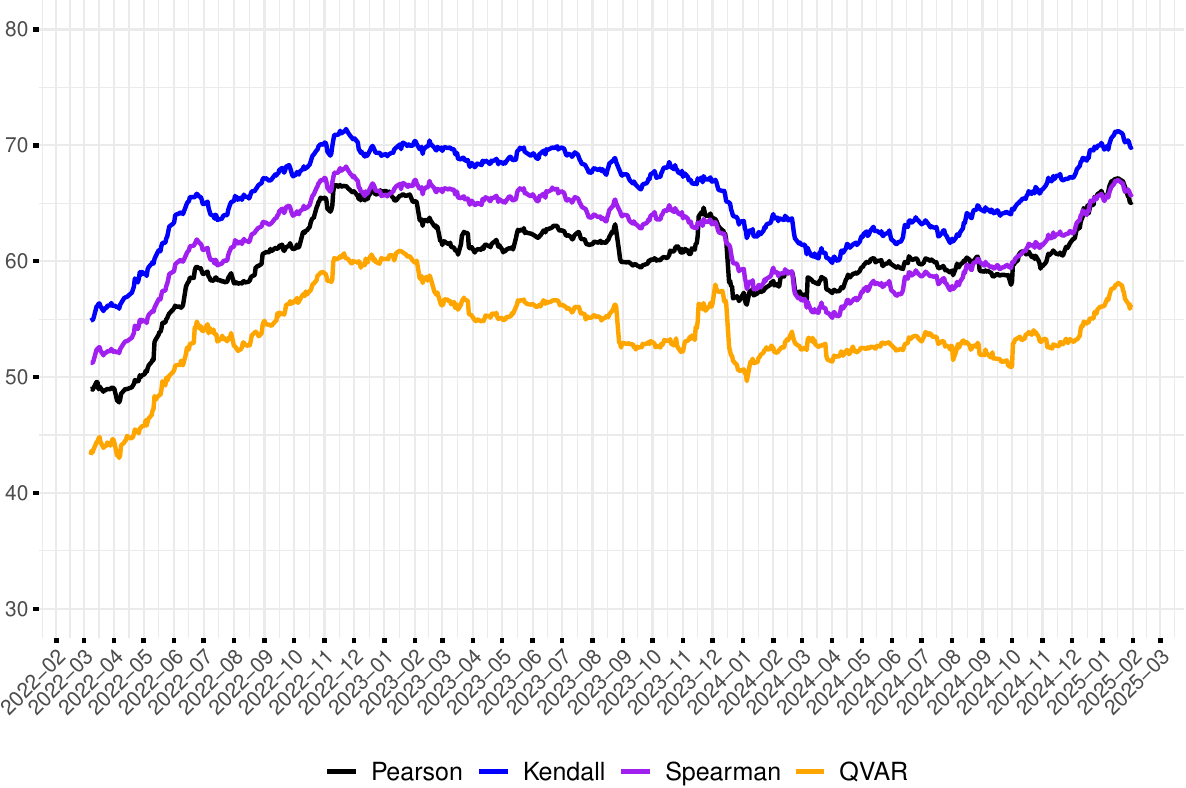}
		\caption{Robustness analysis for TCI.}
		\label{Fig:Robustness}
	\end{figure}

	As shown in  Fig.~\ref{Fig:Robustness}, we conduct a robustness analysis to verify the reliability of our findings, employing multiple approaches, including Pearson, Spearman, and Kendall correlation-based $R^2$ decompositions, as well as quantile VAR (QVAR) \citep{Ando-GreenwoodNimmo-Shin-2022-ManageSci}. All TCIs exhibit a largely consistent pattern, reinforcing the robustness of our findings.
	
	\section{Portfolio and hedging strategies}
	\label{Portfolio}
	\subsection{Bilateral hedge ratios and portfolio weights}
	
	Table~\ref{Tb:HR} presents the summary statistics of bilateral hedge ratios and hedge effectiveness (HE) for AI and green assets. First, AI ETFs and green assets demonstrate relatively strong hedging effectiveness in both long and short positions. Their mutual hedging performance is also robust.  ROBT/SPGB (0.957) and IRBO/SPGB (0.946) show the highest hedge ratios, both statistically significant ($p$-value $<$ 0.01).
	When hedging AI ETFs, clean energy outperforms green bonds in both long and short positions. 
	
	\begin{table}[htb!]
		\centering
		\begin{threeparttable}
			\small
			\relsize{-0.5}
			\caption{Bilateral hedge ratios and hedge effectiveness.}
			\newcolumntype{d}[1]{D{.}{.}{#1}} 
			\label{Tb:HR}
\begin{tabular*}{\textwidth}{@{\extracolsep{\fill}}l d{2.3}  d{1.3}  d{2.3} *{3}{d{1.4}} @{}}
				\toprule
				
				& \multicolumn{1}{c}{\textbf{Mean}} & \multicolumn{1}{c}{\textbf{Std. Dev.}} & \multicolumn{1}{c}{\textbf{5\%}} & \multicolumn{1}{c}{\textbf{95\%}} & \multicolumn{1}{c}{\textbf{HE}} & \multicolumn{1}{c}{\textbf{p-value}} \\
				\midrule
				
				IRBO/FET & 0.007 & 0.011 & -0.010 & 0.022 & 0.018 & 0.886 \\
				IRBO/FIL & 0.008 & 0.017 & -0.018 & 0.041 & 0.018 & 0.855 \\
				IRBO/ICP & 0.000 & 0.017 & -0.024 & 0.037 & 0.014 & 0.887 \\
				IRBO/NEAR & 0.009 & 0.014 & -0.020 & 0.030 & 0.018 & 0.850 \\			
				IRBO/RENDER & 0.012 & 0.010 & -0.002 & 0.029 & 0.020 & 0.840 \\
				IRBO/SPGB & 0.946 & 0.334 & 0.284 & 1.486 & 0.102 & 0.002 \\
				IRBO/SPGTCED & 0.626 & 0.121 & 0.304 & 0.778 & 0.372 & 0.000 \\
				
				ROBT/FET & 0.004 & 0.011 & -0.013 & 0.022 & 0.018 & 0.886 \\
				ROBT/FIL & 0.000 & 0.017 & -0.028 & 0.028 & 0.017 & 0.855 \\
				ROBT/ICP & -0.008 & 0.017 & -0.033 & 0.027 & 0.012 & 0.887 \\
				ROBT/NEAR & 0.004 & 0.013 & -0.017 & 0.024 & 0.018 & 0.850 \\
				ROBT/RENDER & 0.004 & 0.011 & -0.013 & 0.023 & 0.017 & 0.840 \\
				ROBT/SPGB & 0.957 & 0.327 & 0.388 & 1.516 & 0.126 & 0.002 \\		
				ROBT/SPGTCED & 0.609 & 0.114 & 0.367 & 0.776 & 0.412 & 0.000 \\
				
				FET/IRBO & 0.185 & 0.285 & -0.225 & 0.772 & 0.016 & 0.000 \\
				FET/ROBT & 0.106 & 0.291 & -0.316 & 0.571 & 0.015 & 0.000 \\
				FET/SPGB & -0.520 & 1.425 & -3.129 & 2.175 & 0.008 & 0.002 \\
				FET/SPGTCED & -0.104 & 0.284 & -0.570 & 0.326 & 0.009 & 0.000 \\
				
				FIL/IRBO & 0.139 & 0.277 & -0.261 & 0.632 & 0.015 & 0.000 \\
				FIL/ROBT & -0.015 & 0.283 & -0.468 & 0.390 & 0.015 & 0.000 \\
				FIL/SPGB & 0.418 & 0.830 & -0.677 & 1.963 & 0.016 & 0.002 \\
				FIL/SPGTCED & -0.027 & 0.274 & -0.466 & 0.486 & 0.012 & 0.000 \\
				
				ICP/IRBO & 0.035 & 0.302 & -0.421 & 0.713 & 0.013 & 0.000 \\
				ICP/ROBT & -0.120 & 0.306 & -0.527 & 0.543 & 0.011 & 0.000 \\
				ICP/SPGB & -0.165 & 0.715 & -1.623 & 1.093 & 0.003 & 0.002 \\
				ICP/SPGTCED & -0.091 & 0.266 & -0.422 & 0.462 & 0.009 & 0.000 \\

				NEAR/IRBO & 0.115 & 0.375 & -0.880 & 0.624 & 0.018 & 0.000 \\
				NEAR/ROBT & 0.066 & 0.327 & -0.664 & 0.503 & 0.017 & 0.000 \\
				NEAR/SPGB & 0.697 & 0.906 & -1.004 & 2.157 & 0.015 & 0.002 \\
				NEAR/SPGTCED & -0.121 & 0.425 & -1.306 & 0.373 & 0.012 & 0.000 \\
				
				RENDER/IRBO & 0.295 & 0.250 & -0.126 & 0.699 & 0.018 & 0.000 \\
				RENDER/ROBT & 0.108 & 0.331 & -0.400 & 0.777 & 0.015 & 0.000 \\
				RENDER/SPGB & 0.015 & 0.971 & -1.415 & 1.618 & 0.009 & 0.002 \\
				RENDER/SPGTCED & -0.079 & 0.305 & -0.531 & 0.434 & 0.013 & 0.000 \\
				
				SPGB/IRBO & 0.088 & 0.044 & 0.018 & 0.150 & 0.118 & 0.000 \\
				SPGB/ROBT & 0.098 & 0.044 & 0.027 & 0.163 & 0.134 & 0.000 \\
				SPGB/FET & -0.002 & 0.004 & -0.009 & 0.006 & 0.010 & 0.886 \\
				SPGB/FIL & 0.002 & 0.004 & -0.003 & 0.009 & 0.015 & 0.855 \\
				SPGB/ICP & 0.000 & 0.003 & -0.005 & 0.005 & 0.007 & 0.887 \\
				SPGB/NEAR & 0.004 & 0.004 & -0.002 & 0.011 & 0.016 & 0.850\\
				SPGB/RENDER & 0.001 & 0.003 & -0.003 & 0.006 & 0.011 & 0.840 \\
				
				SPGTCED/IRBO & 0.605 & 0.118 & 0.266 & 0.749 & 0.383 & 0.000 \\
				SPGTCED/ROBT & 0.661 & 0.116 & 0.330 & 0.807 & 0.414 & 0.000 \\
				SPGTCED/FET & -0.004 & 0.010 & -0.022 & 0.010 & 0.011 & 0.886 \\
				SPGTCED/FIL & 0.000 & 0.015 & -0.025 & 0.028 & 0.014 & 0.855 \\
				SPGTCED/ICP & -0.009 & 0.016 & -0.036 & 0.020 & 0.008 & 0.887 \\
				SPGTCED/NEAR & -0.003 & 0.014 & -0.034 & 0.018 & 0.011 & 0.850 \\
				SPGTCED/RENDER & -0.002 & 0.011 & -0.020 & 0.016 & 0.013 & 0.840 \\
				\bottomrule
			\end{tabular*}
		\end{threeparttable}
	\end{table}
	
	Second, AI tokens provide limited effectiveness when used to hedge AI ETFs.  
	Hedge ratio means are near zero, HE remains low (ranging from  0.012 to 0.020), and $p$-values exceed 0.8, indicating poor risk reduction capabilities. 
	When tokens are long and AI ETFs are short, hedge ratios increase and HE is still low. The highest hedge ratio is observed in RENDER/IRBO (0.295), while NEAR/IRBO and RENDER/IRBO achieve the highest HE (0.018), still providing minimal risk mitigation. 
	
	Third, green assets fail to provide a strong safe-haven function against cryptos. FIL/SPGB (HE = 0.016, $p$-value $<$ 0.01) shows the best performance, but still offers weak risk reduction.
	When green assets are in a long position, the results lack statistical significance. The HE for all tokens remains no bigger than 0.016 (SPGB/NEAR), indicating the weak hedging ability of AI tokens.
	
	\subsection{Multivariate portfolio analysis}
	
	Table~\ref{Tb:portfolio} show the results for MVP, MCP, MCoP, $\text{MCoP}^C$, and $\text{MCoP}^L$. It is worth emphasizing that MCoP, $\text{MCoP}^C$, and $\text{MCoP}^L$ are derived from total, contemporaneous, and lagged connectedness, respectively.
	We first discuss the performance of AI ETFs in the MVP strategy. The relatively high HE values (IRBO: 0.905, ROBT: 0.890) indicate that multivariate portfolios are effective in reducing the risk of AI ETFs. Even though AI ETFs are the primary transmitters of shocks in the return transmission analysis, their weights remain relatively low (IRBO: 2.4\%, ROBT: 3.8\%). This suggests that their role in the MVP portfolio is limited. 
	In the MCP, the allocation to AI ETFs increases significantly (IRBO: 6.0\%, ROBT: 12.5\%), indicating an enhanced contribution to the portfolio. However, their HE values turn negative (IRBO: $-2.215$, ROBT: $-2.728$), indicating that this portfolio fails to effectively mitigate volatility in AI ETFs and may even increase risk. 
	In the MCoP method, the allocation to AI ETFs further increases (IRBO: 10.5\%, ROBT: 11.5\%), reinforcing their role within the portfolio. Nevertheless, their HE values continue to decline (IRBO: $-3.375$, ROBT: $-4.073$). Furthermore, the MCoP , $\text{MCoP}^C$, and $\text{MCoP}^L$  allocation methods assign very close weights to AI and green assets.
	Overall, despite Figs.~\ref{Fig:Static:Connectedness}, \ref{Fig:NET} and \ref{Fig:Network}  indicating that AI ETFs are major risk spillovers, their weights in these portfolios remain relatively limited.

	Next, we analyze the role of AI Tokens across different portfolio strategies. In the MVP, AI tokens (FET, FIL, ICP, NEAR, RENDER) have minimal weights, each below 1\%, limiting their overall impact on the portfolio. 	
	Furthermore, their HE values exceed 0.99, suggesting that this investment strategy effectively minimizes their risk.
	In the MCP portfolio, despite a decrease in HE values (ranging between 0.783 and 0.854), the weights of AI Tokens increase significantly (e.g., RENDER: 12.6\%, FET: 11.6\%), indicating their enhanced role within the portfolio. 
	In the MCoP, the weights of AI Tokens further increase, reflecting their rising importance in this investment strategy. Their HE values remain above 0.7 (ranging between 0.705 and 0.802), suggesting that this strategy  contribute to volatility reduction of AI tokens, though to a lesser extent than in the MVP and MCP. Moreover, portfolios constructed using $\text{MCoP}^C$ and $\text{MCoP}^L$ also exhibit a similar pattern. 
	Overall, AI Tokens exhibit higher allocation ratios for both MCP and MCoPs, whereas their weights in the MVP remains low due to higher volatility, limiting their overall influence.
	
	Finally, we examine the performance of green assets in the portfolios. Green assets remain the core holdings in the MVP, with the highest allocation. SPGB holds an overwhelming weight of 91.2\%, while SPGTCED accounts for 0.7\%, likely reflecting their respective roles as a risk receiver and a risk contributor.	Accordingly, SPGTCED has an HE value of 0.884, demonstrating a significantly stronger risk reduction effect compared to SPGB, which has a negative HE value of $-$0.05.		
	From the MCP procedure, the weight of green assets declines significantly (SPGB: 19.7\%, SPGTCED: 13.2\%, total: 32.9\%), and their HE values continue to decline, indicating a weakening risk reduction in this portfolio. 
	In the MCoP, the allocation to green assets decreases further (SPGB: 11.2\%, SPGTCED: 11.0\%, total: 22.2\%), with HE values continuing to drop, suggesting a further reduction in their role. 	
	Despite \Cref{Fig:Static:Connectedness,Fig:NET,Fig:Network}  demonstrating that clean energy transmits shocks to other markets, its weight in the portfolio is not dominant. Again, the results from $\text{MCoP}^C$ and $\text{MCoP}^L$ are very close to those of $\text{MCoP}$.	
	This finding aligns with previous literature \citep{Tiwari-Abakah-Gabauer-Dwumfour-2022-GlobFinancJ}, which also found that while clean energy is a risk contributor, its allocation in investment portfolios is not necessarily the highest, whereas green bonds tend to account for a larger proportion of portfolio allocation.

	Interestingly, in the case of the MVP, asset allocations, except for AI tokens, seemly correspond to their roles. Risk receivers have relatively higher weights, while risk transmitters have lower weights. The results of MCP and MCoPs are quite similar, possibly because both focus on reducing correlations across assets. All the portfolios reduce the risk of AI tokens but increase volatility of the green bond.
	
	\subsection{Portfolio performance}
	
\begin{figure}[htp]
		\centering
		\includegraphics[width=0.5\linewidth]{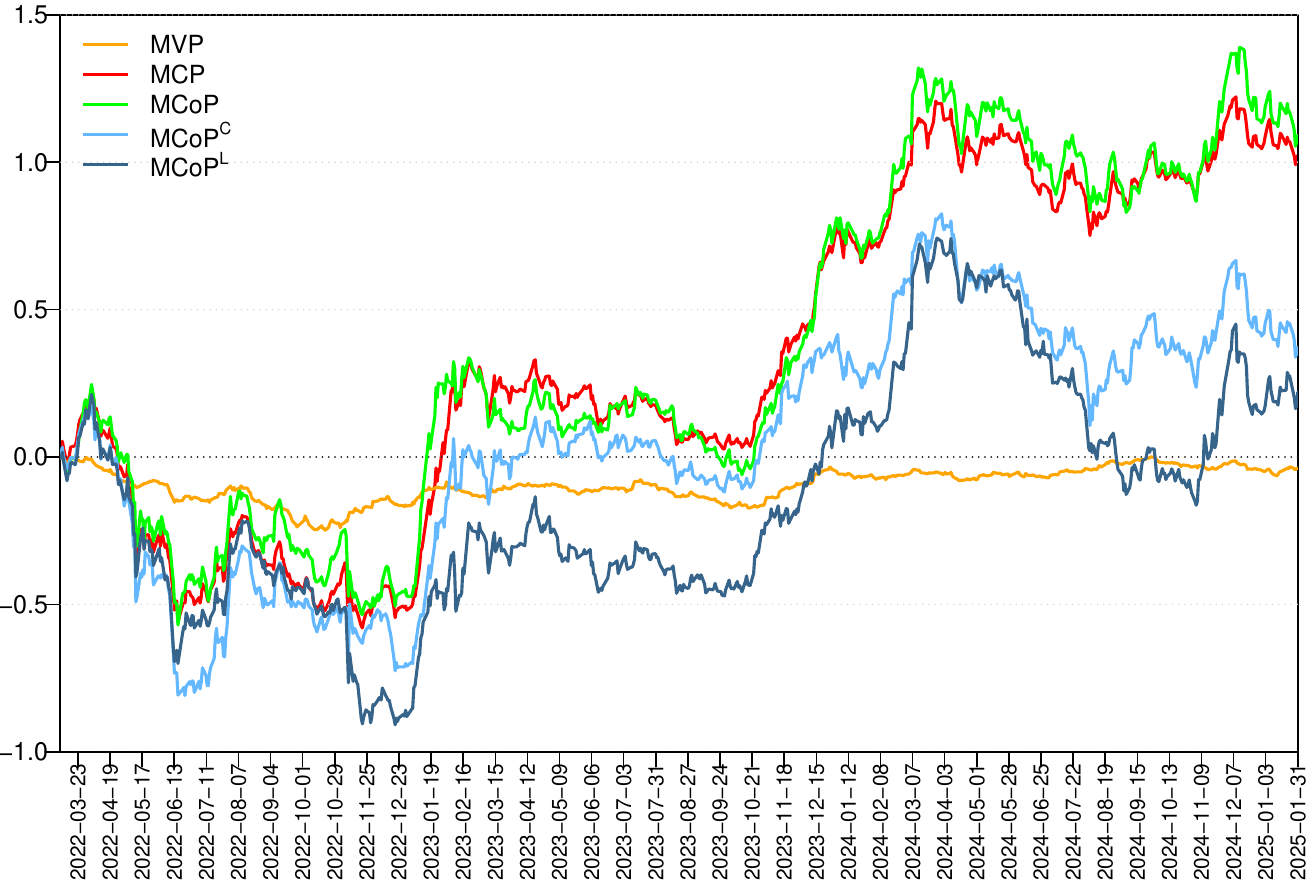}
		\caption{Cumulative returns of different portfolios.}
		\label{Fig:cumsumR}
	\end{figure}

	We illustrate the cumulative portfolio returns in Fig.~\ref{Fig:cumsumR}. This figure presents the cumulative returns of portfolio strategies. MVP remains relatively stable with minimal fluctuations, indicating a low-risk but low-return profile, while MCP and MCoPs show significant growth, especially from early 2023 onward. In the early period, all portfolios experience negative returns, but MCP and MCoPs recover strongly, whereas MVP remains around the zero level. MCP and MCoP follow similar trends, with MCoP generally achieving slightly higher peaks, suggesting better performance in specific market conditions. Other studies on investment strategy comparisons also documents negative cumulative portfolio returns \citep{Rubbaniy-Khalid-Syriopoulos-Polyzos-2024-JFuturesMark,Lang-Hu-Corbet-Goodell-2023-FinancResLett}.
	
		\begin{table}[htp]
		\centering
		\caption{Multivariate portfolio weights.}
		\label{Tb:portfolio}				
			\newcolumntype{d}[1]{D{.}{.}{#1}} 
	\begin{tabular*}{\textwidth}{@{\extracolsep{\fill}} c l   *{4}{d{1.3}} d{2.3} d{1.3} @{}}
		\toprule		
		&	& \multicolumn{1}{c}{\textbf{Mean}} & \multicolumn{1}{c}{\textbf{Std.Dev.}} & \multicolumn{1}{c}{\textbf{5\%}} & \multicolumn{1}{c}{\textbf{95\%}} & \multicolumn{1}{c}{\textbf{HE}} & \multicolumn{1}{c}{\textbf{p-value}} \\
			\midrule
			MVP&IRBO & 0.024 & 0.028 & 0.000 & 0.072 & 0.905 & 0.000 \\
			&ROBT & 0.038 & 0.044 & 0.000 & 0.121 & 0.890 & 0.000 \\
			&FET & 0.007 & 0.006 & 0.000 & 0.016 & 0.996 & 0.000 \\
			&FIL & 0.003 & 0.003 & 0.000 & 0.009 & 0.994 & 0.000 \\
			&ICP & 0.006 & 0.005 & 0.000 & 0.015 & 0.994 & 0.000 \\
			&NEAR & 0.000 & 0.001 & 0.000 & 0.000 & 0.994 & 0.000 \\
			&RENDER & 0.002 & 0.002 & 0.000 & 0.007 & 0.996 & 0.000 \\
			&SPGB & 0.912 & 0.037 & 0.846 & 0.970 & -0.050 & 0.508 \\
			&SPGTCED & 0.007 & 0.014 & 0.000 & 0.043 & 0.884 & 0.000 \\
			
			\midrule
			MCP&IRBO & 0.060 & 0.082 & 0.000 & 0.240 & -2.215 & 0.000 \\
			&ROBT & 0.125 & 0.087 & 0.000 & 0.232 & -2.728 & 0.000 \\
			&FET & 0.116 & 0.047 & 0.028 & 0.183 & 0.854 & 0.000 \\
			&FIL & 0.080 & 0.051 & 0.007 & 0.170 & 0.792 & 0.000 \\
			&ICP & 0.113 & 0.064 & 0.000 & 0.207 & 0.783 & 0.000 \\
			&NEAR & 0.052 & 0.042 & 0.000 & 0.124 & 0.785 & 0.000 \\
			&RENDER & 0.126 & 0.027 & 0.086 & 0.170 & 0.851 & 0.000 \\
			&SPGB & 0.197 & 0.034 & 0.139 & 0.261 & -34.611 & 0.000 \\
			&	SPGTCED & 0.132 & 0.034 & 0.061 & 0.176 & -2.927 & 0.000 \\
			\midrule
			
			MCoP&IRBO & 0.105 & 0.112 & 0.000 & 0.250 & -3.375 & 0.000 \\
			&ROBT & 0.115 & 0.120 & 0.000 & 0.250 & -4.073 & 0.000 \\
			&FET & 0.104 & 0.115 & 0.000 & 0.250 & 0.802 & 0.000 \\
			&FIL & 0.109 & 0.111 & 0.000 & 0.250 & 0.718 & 0.000 \\
			&ICP & 0.114 & 0.119 & 0.000 & 0.250 & 0.705 & 0.000 \\
			&NEAR & 0.111 & 0.116 & 0.000 & 0.250 & 0.708 & 0.000 \\
			&RENDER & 0.120 & 0.128 & 0.000 & 0.333 & 0.798 & 0.000 \\	
			&SPGB & 0.112 & 0.118 & 0.000 & 0.250 & -47.457 & 0.000 \\
			&SPGTCED & 0.110 & 0.113 & 0.000 & 0.250 & -4.343 & 0.000 \\
			\midrule
			
			MCoP$^C$&IRBO & 0.116 & 0.115 & 0.000 & 0.250 & -3.426 & 0.000 \\
			&ROBT & 0.105 & 0.116 & 0.000 & 0.250 & -4.132 & 0.000 \\
			&FET & 0.112 & 0.119 & 0.000 & 0.250 & 0.800 & 0.000 \\		
			&FIL & 0.104 & 0.111 & 0.000 & 0.250 & 0.714 & 0.000 \\
			&ICP & 0.113 & 0.115 & 0.000 & 0.250 & 0.701 & 0.000 \\
			&NEAR & 0.109 & 0.119 & 0.000 & 0.250 & 0.704 & 0.000 \\
			&	RENDER & 0.113 & 0.111 & 0.000 & 0.250 & 0.795 & 0.000 \\
			&SPGB & 0.115 & 0.116 & 0.000 & 0.250 & -48.025 & 0.000 \\
			&SPGTCED & 0.113 & 0.120 & 0.000 & 0.250 & -4.406 & 0.000 \\
			\midrule
			
			MCoP$^L$&	IRBO & 0.105 & 0.113 & 0.000 & 0.250 & -3.753 & 0.000 \\
			&ROBT & 0.121 & 0.124 & 0.000 & 0.333 & -4.512 & 0.000 \\
			&FET & 0.109 & 0.114 & 0.000 & 0.250 & 0.785 & 0.000 \\
			&FIL & 0.106 & 0.110 & 0.000 & 0.250 & 0.693 & 0.000 \\
			&ICP & 0.106 & 0.119 & 0.000 & 0.250 & 0.679 & 0.000 \\
			&	NEAR & 0.118 & 0.113 & 0.000 & 0.250 & 0.682 & 0.000 \\
			&	RENDER & 0.112 & 0.114 & 0.000 & 0.250 & 0.780 & 0.000 \\
			&	SPGB & 0.116 & 0.111 & 0.000 & 0.250 & -51.654 & 0.000 \\
			&	SPGTCED & 0.108 & 0.116 & 0.000 & 0.250 & -4.806 & 0.000 \\
			\bottomrule
		\end{tabular*}
	\end{table}

	Finally, the Sharpe ratios of the three portfolios are shown in Table~\ref{Tb:portfolioperformance}. The MCP emerges as the most optimal strategy, as it achieves the highest return and Sharpe ratios across all evaluated metrics. While the MVP effectively minimizes volatility, its negative return and Sharpe ratios render it an unattractive investment option. $\text{MCoP}^C$, and $\text{MCoP}^L$ have negative return and Sharpe ratios. The MCoP offers a moderate risk-return profile, yet it fails to surpass MCP. Therefore, for investors seeking an optimal balance between risk and return, the MCP portfolio represents the most advantageous selection. This finding is similar with 
	\cite{Ziadat-Mensi-AlKharusi-Vo-Kang-2024-EnergyEcon}, who also observe a negative Sharpe ratio.

\begin{table}[htp]
	\centering
	\caption{Portfolio performance.}
	\label{Tb:portfolioperformance}
\newcolumntype{d}[1]{D{.}{.}{#1}} 
\begin{tabular*}{\textwidth}{@{\extracolsep{\fill}}l d{2.4}  *{2}{d{1.4}} d{2.4} d{2.5} @{}}
		\toprule
		& \multicolumn{1}{c}{\textbf{MVP}} & \multicolumn{1}{c}{\textbf{MCP}} & \multicolumn{1}{c}{\textbf{MCoP}}&
		\multicolumn{1}{c}{\textbf{{MCoP}$^{C}$}}& \multicolumn{1}{c}{\textbf{{MCoP}$^{L}$}}\\
		\midrule
Return & -0.0165 & 0.2664 & 0.2438 & -0.0335 & -0.0932\\
StdDev & 0.0835 & 0.4860 & 0.5669 & 0.5702 & 0.5910\\
Sharpe Ratio (StdDev) & -0.1978 & 0.5482 & 0.4300 & -0.0588 & -0.1578\\
Sharpe Ratio (VaR) & -1.9865 & 5.8478 & 4.8230 & -0.6228 & -1.7048\\
Sharpe Ratio (CVaR) & -1.4889 & 4.2027 & 3.8360 & -0.4194 & -1.2778\\
		\bottomrule
	\end{tabular*}
\end{table}

	\section{Conclusion}
	\label{Conclusion}
		
	This study employs the $R^{2}$ decomposition to examine the risk spillover dynamics among AI ETFs, AI tokens, and green assets over the period from May 25, 2021, to January 31, 2025. Our findings reveal that contemporaneous spillovers dominate the total connectedness index, while lagged spillovers remain relatively stable. Furthermore, AI ETFs and clean energy emerge as risk transmitters, whereas AI tokens and green bonds primarily act as risk receivers.
	In addition, AI tokens are challenging to hedge and offer little benefit in hedging other assets when considering bilateral hedge ratios.
	However, clean energy and AI ETFs exhibit stronger hedging effectiveness regardless of the long or short position taken.
	Additionally, we analyze the performance of the following portfolio strategies: MVP, MCP, and MCoPs. The results show that while all the approaches reduce volatility of AI tokens significantly, they increase the risk associated with green bonds. However, only the MVP decreases the risk of AI ETFs and clean energy, while the other strategies actually increase their risk. Finally, MCP exhibits superior performance than the other portfolios. In this investment strategy, green bonds have the largest allocation, while IRBO has the smallest.
	
	The nature of these assets as financial instruments might be helpful to better understand their risk transmission. Although AI tokens are thematically linked to artificial intelligence, they fundamentally belong to the category of crypto assets, which are often speculative, volatile, and sensitive to external shocks \citep{Kukacka-Kristoufek-2023-FinancInnov,Osman-Galariotis-Guesmi-Hamdi-Naoui-2024-IntRevEconFinanc}. 
	Their highly synchronized price movements suggest internal spillovers and clustering effects.

    On the contrary, AI ETFs represent firms with actual research and development capabilities in AI. The pricing of these assets is more directly driven by market expectations regarding AI advancement and earnings potential. Spillovers from AI ETFs to green assets may partly reflect market perceptions of AI's enabling role in advancing environmental sustainability, as suggested by existing literature \citep{Vinuesa-Azizpour-Leite-Balaam-Dignum-Domisch-Fellander-Langhans-Tegmark-FusoNerini-2020-NatCommun, Entezari-Aslani-Zahedi-Noorollahi-2023-EnergyStrategRev, Nahar-2024-TechnolForecastSocChang}.
    
    Green bonds, as relatively stable and policy-backed debt instruments, act as risk receivers within the financial system, consistent with findings of \cite{Tiwari-Abakah-Gabauer-Dwumfour-2022-GlobFinancJ}. In contrast, clean energy equities are characterized by technological orientation and strong growth potential. \cite{Zeng-Abedin-Zhou-Lu-2024-IntRevFinancAnal} and \cite{Tiwari-Abakah-Gabauer-Dwumfour-2022-GlobFinancJ} also indicate that they may serve as sources of risk spillovers in the system. Such spillovers may, at least in part, reflect the broader influence of clean energy technologies. Therefore, future research could empirically examine how asset-specific characteristics and underlying structures influence the risk spillover across different types of assets.
	
	Our work has the following implications: First, given the risk-spreading nature of AI ETFs and clean energy assets, regulators should closely monitor their volatility and systemic impact. Implementing safeguards to prevent excessive market fluctuations is important. Additionally, policymakers should pay close attention to the interconnectedness between technology-related financial instruments and green assets, particularly within the frameworks of sustainable finance and climate risk management. In this context, enhanced cross-market risk monitoring and response mechanisms are essential. Moreover, aligning technology policy with green finance policy may be necessary to effectively address the systemic implications of such interlinkages.
		
	Second,  investors should consider the high spillover risks associated with AI ETFs and clean energy. Although the two assets classes are highly popular, their risk transmission role suggests that they should be complemented with strong hedging assets. 	
	Besides that, given the ineffective hedging of AI tokens with other assets, investors should closely monitor their exposure to AI-related cryptos and adjust their allocations accordingly. MCP outperforms MVP and MCoPs, making it the most suitable strategy for investors, especially those interested in AI tokens. Investors should also be mindful of potential herding behavior among AI tokens, as their highly synchronized dynamics and strong interconnections may pose hidden systemic risks.

\end{document}